\documentclass[smallextended]{svjour3}
\usepackage{graphicx,epsf, epsfig, amssymb}
\usepackage[usenames]{color}

\journalname{GRG}

\def\prd{Phys.~Rev.~D }       
\def\cqg{Class.~Quant.~Grav. }

\newcommand{\be}{\begin{equation}}
\newcommand{\ee}{\end{equation}}
\newcommand{\bel}[1]{\begin{equation}\label{#1}}
\newcommand{\ba}{\begin{eqnarray}}
\newcommand{\ea}{\end{eqnarray}}
\newcommand{\bal}[1]{\begin{eqnarray}\label{#1}}

\begin{document}

\title{Data Analysis Challenges for the Einstein Telescope}
\author{Leone Bosi \and Edward K. Porter}\institute{L. Bosi \at INFN Perugia, via Pascoli I-6123, Perugia, Italy\\ \\ E.~K.~Porter \at APC, UMR 7164, Universit\'e Paris 7 Denis Diderot, 10, rue Alice Domon et L\'{e}onie Duquet, 75205 Paris Cedex 13, France}

\date{July 26, 2009}

\maketitle
\begin{abstract}
The Einstein Telescope is a proposed third generation gravitational wave detector that will operate in the region of 1 Hz to a few kHz.  As well as the inspiral of compact binaries composed of neutron stars or black holes, the lower frequency cut-off of the detector will open the window to a number of new sources.  These will include the end stage of inspirals, plus merger and ringdown of intermediate mass black holes, where the masses of the component bodies are on the order of a few hundred solar masses.  There is also the possibility of observing intermediate mass ratio inspirals, where a stellar mass compact object inspirals into a black hole which is a few hundred to a few thousand times more massive.  In this article, we investigate some of the data analysis challenges for the Einstein Telescope such as the effects of increased source number, the need for more accurate waveform models and the some of the computational issues that a data analysis strategy might face.
\end{abstract}

\section{Introduction} 
The Einstein Telescope (ET)~\cite{et} is a proposed third generation gravitational wave (GW) detector which will operate in a frequency range down to $\sim1$ Hz.  This detector will compliment the current range of ground based detectors such as LIGO~\cite{ligo} and VIRGO~\cite{virgo} which have lower frequency cut-offs of $\sim40$ and $\sim30$ Hz respectively, and their second generation versions which should operate with lower frequency cut-offs of $\sim10$ Hz~\cite{ligo2,virgo2}.  ET will also be important as it will fill some of the gap between the planned Laser Interferometer Space Antenna (LISA)~\cite{lisa}, which will have an upper frequency cut-off of approximately 0.1 Hz,  and the ground based interferometers.

Many of the sources observable with ET are similar to those that should be visible with the current ground based detectors, i.e. the inspiral of compact objects such as neutron stars and black holes.  However, the main difference between ET and current detectors will be an order of magnitude increase in sensitivity, with a low frequency cut-off of 1-2 Hz.  Because of this,  the signals will be observable for much longer, with some sources being observable up to a day in length.  As well as the common sources, there are other sources that will be new to the ground based detectors.  These sources will include the inspiral of intermediate mass black hole binaries (IMBHBs), i.e. comparable mass systems with masses from a few hundred to a few thousand solar masses, and  possibly intermediate mass ratio inspirals (IMRIs), i.e. systems composed of a black hole of a few hundred to a few thousand solar masses and a stellar mass compact object.

Another difference between ET and current detectors is that we would expect to have a large number of sources, with quite high signal to noise ratios (possibly as high as a hundred times higher than current capabilities).  While the high source count is desirable, it is possible that at low frequencies it could lead to a source of confusion noise from a high number of neutron star binaries all inhabiting the same frequency space at lower frequencies.

The longer duration signals, combined with the possibilities of high signal to noise ratios and possibly even source confusion will present a challenge to the data analysis community.  The detection and estimation of parameters will require sophisticated algorithms that take both time and computational constraints into account, as our purpose is to carry out GW astronomy.

The outline of the article is as follows : in Section~\ref{gba} we describe some of the algorithms that are currently used in the detection of both burst and inspiral sources in the ground-based community.  Section~\ref{ccg} contains a discussion of the computational cost related to an inspiral search.  In Section~\ref{ets} we look in more detail at some of the sources that should be visible to ET and how they may effect the data analysis strategies.  In Section~\ref{lisaal} we discuss the advantage of using LISA oriented algorithms to search in the higher dimensional search spaces that will be open to ET.  Finally, in Section~\ref{etcc} we investigate some of the computational issues that will be important for ET.

\section{Status of current ground based algorithms \label{gba}}
In this section we provide a description of the search algorithms currently used in the actual inspiral and burst analysis procedure. 
\subsection{Burst signals}
Burst signals have typical duration less than a second due to transients of GW radiation. Theoretical models for this class of signals are usually not very accurate, due to a lack of theory on the complexity of the production processes, that further depend on a high number of physical parameters. This fact does not allow us to apply the standard matched filtering procedure (which we will describe later).  Due to the sensibility on the model parameters and accuracy, time/frequency energy filters are mainly applied instead. These detection algorithms look for excess power and are characterized by having a reduced detection efficiency with respect to the matched filtering method, but by having a lower rejection criterion, it allows one to detect signals that are not very well modelled.

Possible burst sources are stellar core collapse, pulsar glitches, compact BH or NS binary systems, r-modes on the stellar surface and any other events that can produce a shock in the GW field.  The output signals of the actual detectors are not stationary in terms of the effect of transients introduced by instrumental and environmental sources.  Due to this,  the set of filters for burst detection often mistake such noise structures as GW signals, producing a number of triggers related to false alarms.

A first approach to discriminate between triggers related to the noise, and triggers related to a hypothetical signal, is the detector signal characterization. This activity allows scientists to discover, analyze and characterize the sources of each spurious noise event and use this information to veto associated triggers. However, even using this approach, the number of surviving triggers is too high to provide a good detection confidence with a single detector analysis.

This is the main reason why a multi-detector analysis is heavily used.  This permits us to introduce a further constraint : triggers identified in each individual detector have to be coincident with triggers in other detectors in a defined time and/or frequency window. The strategy followed and the size of the windows used depend on the signal models, algorithms and the time signal delay between different detectors. Another advantage of using such an approach, with at least three detectors, is the possibility of locating the source in the sky.

\begin{figure}[t]
\centering
\epsfig{file=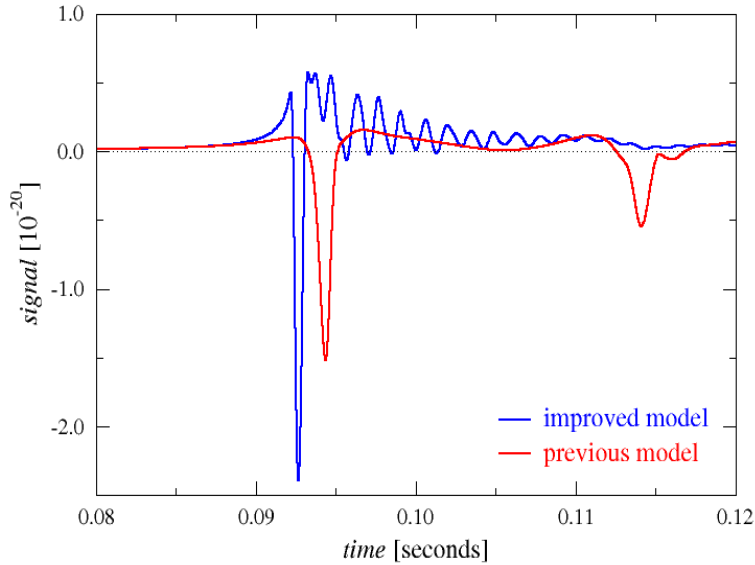, width=4in, height=3.in}
\caption{A comparison of the waveform generated using old and new models of GW emission from a core-collapse supernova.}
\label{fig:burstExample}
\end{figure}

As previously stated, the theoretical models for burst sources are not very advanced.  For this very reason, burst searches employ generic methods, making some minimal assumptions on the time-frequency signature of the expected GW signal. In the case of having accurate theoretical predictions, the models can still be very prohibitive for many potential sources due to the high number of unknown parameters.  In this case, we can use information from how some of the waveforms should look like.   In figure \ref{fig:burstExample}, we show an example of a new and old GW burst model produced by core-collapse supernovae.

Here we give a brief report on some detection algorithms used in burst detection : 

\begin{itemize}
\item Zwerger-Muller (\cite{burstZM1}\cite{burstZM2}) - used to identify supernovae waveforms as well as dumped sinusoids from black hole ringdowns.
\item BlockNormal (\cite{burstBN}) - identifies change points of the time series by monitoring the mean and the variance of the time series used. 
\item Excess Power (\cite{burstEP}) - an excess power method, designed to detect short-duration ($\leq$0.5 s) burst signals of unknown waveform, such as those from supernovae or black hole mergers. 
\item WaveBurst (\cite{burstWB}) -  a time-frequency method looking for excess power, based on a signal decomposed in the wavelet domain.
\item Multiresolution (\cite{burstQP}) -  implementation of a logarithmic tiling of the time–frequency plane in order to efficiently detect astrophysically unmodelled bursts of gravitational radiation.
\item Exponential Gaussian Correlator (\cite{burstEGC}) - an algorithm based on a matched filter using exponential Gaussian templates.
\end{itemize}

There are many other methods investigate and employed in gravitational wave burst analysis. Usually in real applications, the Receiver Operative Characteristic (ROC)~\cite{KAY} for each method is estimated, using a simulated signal.  The ROC is a tool that allows one to discriminate between optimal and suboptimal detectors by evaluating detection and false alarm probability on a changing threshold.  This is done in order to evaluate the detection algorithm efficiency and to chose a subset of the above techniques for the final data analysis.  Moreover, procedures and algorithms have been evolved to perform a cross-correlation between detectors,  taking into account the antenna response, noise and data quality of each detector\cite{Burst2}.

Obviously, here we are citing only the concept at the base of burst detection.  The true GW burst detection procedure is much more involved. For a deeper knowledge on this topic  there is a lot of literature and some important references can be found in \cite{Burst2,Burst1}.

\subsection{inspiral signals}

One of the most important sources of GWs will be the inspiral, merger and ringdown of compact binary sources.   The component bodies of the binary system lose orbital energy due to the emission of GWs.  The dimension of the inspiral effect, the GW strain amplitude,  depends strongly on the masses of the two bodies. In particular, sources that can produce an inspiral detectable with our instruments need to be composed of extremely compact objects.  For ground based detectors, the main systems of interest are composed of neutron star - neutron star (NS-NS), neutron star - black hole (NS-BH) and black hole - black hole (BH-BH) binaries. 

With the current detectors, an inspiral is detectable if the binary is visible in the frequency range where the detectors have the best sensitivity. This typical boundary is between 24-40 Hz up to 1000-2000 kHz, which is associated with the end of the inspiral, and the beginning of the merging phase and ringdown. If the binary system is close enough, and hence detectable, a standard detector like VIRGO or LIGO only has on the order of seconds to a few minutes to observe this signal.  

Unlike the burst case, good theoretical models exist for inspiralling compact objects\cite{biww}. An inspiral signal depends on a set of parameters, such as the component stars masses, inclination angle, polarization angle, arrival time, amplitude, etc... that are all unknown.   As the waveforms, and thus the phase of the waves, are well modelled, the standard detection algorithm is matched filtering. This method is based on correlating known theoretical waveforms, called templates, with the detector output in order to detect the presence of the unknown signal in the data stream. It is possible to demonstrate that, under the hypothesis of additive stochastic noise, the matched filter is the optimal linear filter for signal-to-noise ratio maximization\cite{KAY}.

The condition in which the matched filter operates for inspiral GW detection is complex, due to the number of unknown inspiral parameters.  In fact we also do not know when a signal will arrive, from where in the sky and from which source-type. The only way to conduct a search is to create a bank of templates, covering the space of parameters in a proper manner, and compare each of these reference signals with the detector stream via matched filtering. 

This kind of detection procedure is computionally demanding, as the matched filter involves a correlation, which performed in the time domain has a complexity $O(N^2)$, where $N$ is the vector length. This operation is usually made in frequency domain, because, thanks to the FFT algorithm, the complexity is reduced to $O(N\log (N))$. This is much better, but still computationally heavy. Moreover, for each detector stream time-slice, we have to apply the full template bank.  This means roughly, that if we have  $M$ templates, the complexity of the problem increase linearly with the template bank size,  $O(M N\log (N))$. Currently the typical size of a template bank is of the order of some thousands, depending on the low frequency cut-off frequency of detector sensitivity. 

Obviously, in reality, the situation in much more complex, involving the distribution of processes across more computing nodes, retrieving and compacting filter outputs with a clusterization procedure (i.e., a way of reducing a number of triggers that may be associated with the same time window, to a single trigger based on a number of criterea).   Moreover, as for the burst case, even if the matched filter is more selective, there are still many false alarm triggers related to noise artifacts. For this, veto procedures have been defined based on detector characterization, data quality and a time/frequency inspiral consistency veto~\cite{chi2}. This last procedure is based on a consistency check that verifies  the  time/frequency distribution of the observed event with the expected event. This filter which is called $\chi^2$, and its derivative, require an order of magnitude increase in computational power as compared to the detection procedure, because essentially we perform several matched filters in ten or more sub-frequency ranges.  Also in the case of inspiral detection, the multi-detector analysis is heavily used introducing further constraints, as with the burst case.

\section{Detection algorithms and computational cost\label{ccg}}
\subsection{Inspiral analysis}
As reported in the previous section, matched filtering is the base search algorithm involved in the inspiral search analysis. This algorithm demands high computional power and depends exponentially on the low frequency detector cut-off from which the analysis starts. This choice also defines the detector sensitivity at the low frequency cut-off.

For example, if we consider LIGO detector data, the analysis starts at 40 Hz, while if we consider VIRGO data, due to its optimal performance at low frequency, the analysis begins at 30 Hz. This 10 Hz difference produces a large increase in template bank size. In fact if we consider a minimal match (a measure of the correlation between a template and potential signal) of $95\%$ in the template bank spacing, we observe that in the LIGO case we need roughly $2000$ templates to cover the individual mass parameter space of $[1-3]\,M_{\odot}$, while in the Virgo case we need roughly $7000$ templates. The situation can be much worse if we consider a minimal match of $98\%$, in this case the number of template for Virgo increases to $15000$. There are several developed pipelines involved in the detection of compact binaries that are designed to speed up computational time using different approaches.  A brief description can be found in~\cite{InspiralPipeline1,InspiralPipeline2,InspiralPipeline3,InspiralPipeline4}.

If we consider the Virgo data, starting from 30hz,  adding the time processing constraint imposed by the online analysis, we can estimate the computational requirements to perform the matched filtering technique over the whole template bank. The number of required computational nodes (class opteron 2.2GHz) is roughly 40. Usually the coalescing binaries analysis involves other algorithms, such as chi-squared time-frequency veto procedure~\cite{chi2} that can increase the computational requirements by an order of magnitude . 

The real analysis pipelines are composed of many steps, introducing many layers, such as a hierarchical procedure in order to optimize computation. The network analysis puts constraints in the time/physical parameters coincidences, detectors data quality, parameters reconstruction etc.  Parameter reconstruction of a GW, where it can be parametrized , usually involve parameters of the binary system whose reconstructed values are not independent.   One way to obtain the posterior probability distribution of the various parameters is by use of Markov Chain Monte Carlo~\cite{mcmc1,mcmc2} method.  We will cover this method in a little more detail later on.

The coherent network data-analysis strategy is a very demanding strategy. In this case the detector network is seen as a single detector, and all the statistic are redefined in this context.  The dimensionality of the parameter space increases due to extra parameters that are now resolvable due to the fact that we are using a network, and with that, the number of templates needed increases exponentially. As reported in \cite{InspiralCh}, this analysis requires tens of Tflops  for a three-detector network.

\begin{figure}[t]
  \vspace{5pt}

  \centerline{\hbox{\hspace{0.0in} 
    \epsfxsize=2.6in
   \epsffile{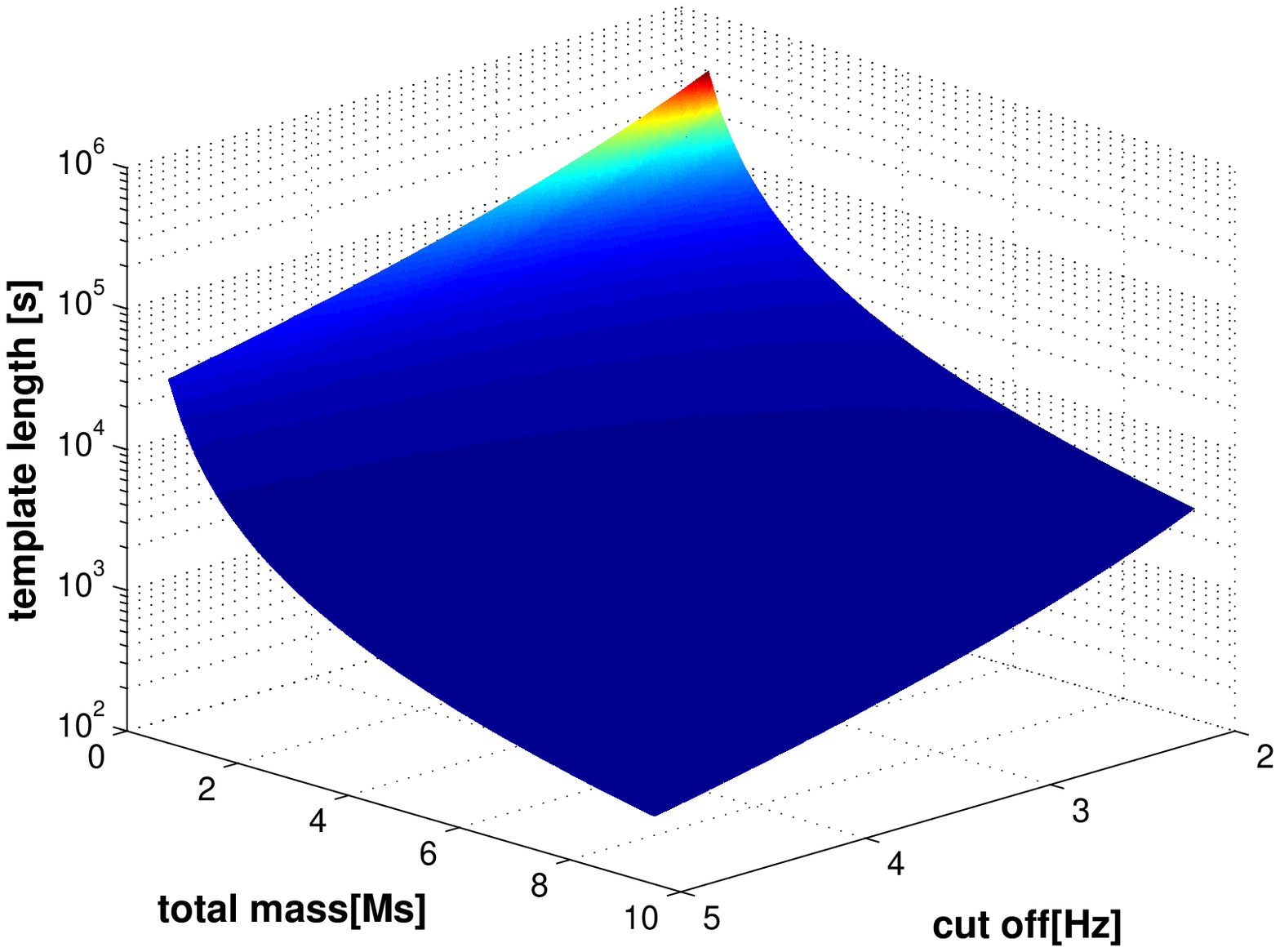}
    \hspace{0.05cm}
    \epsfxsize=2.7in
    \epsffile{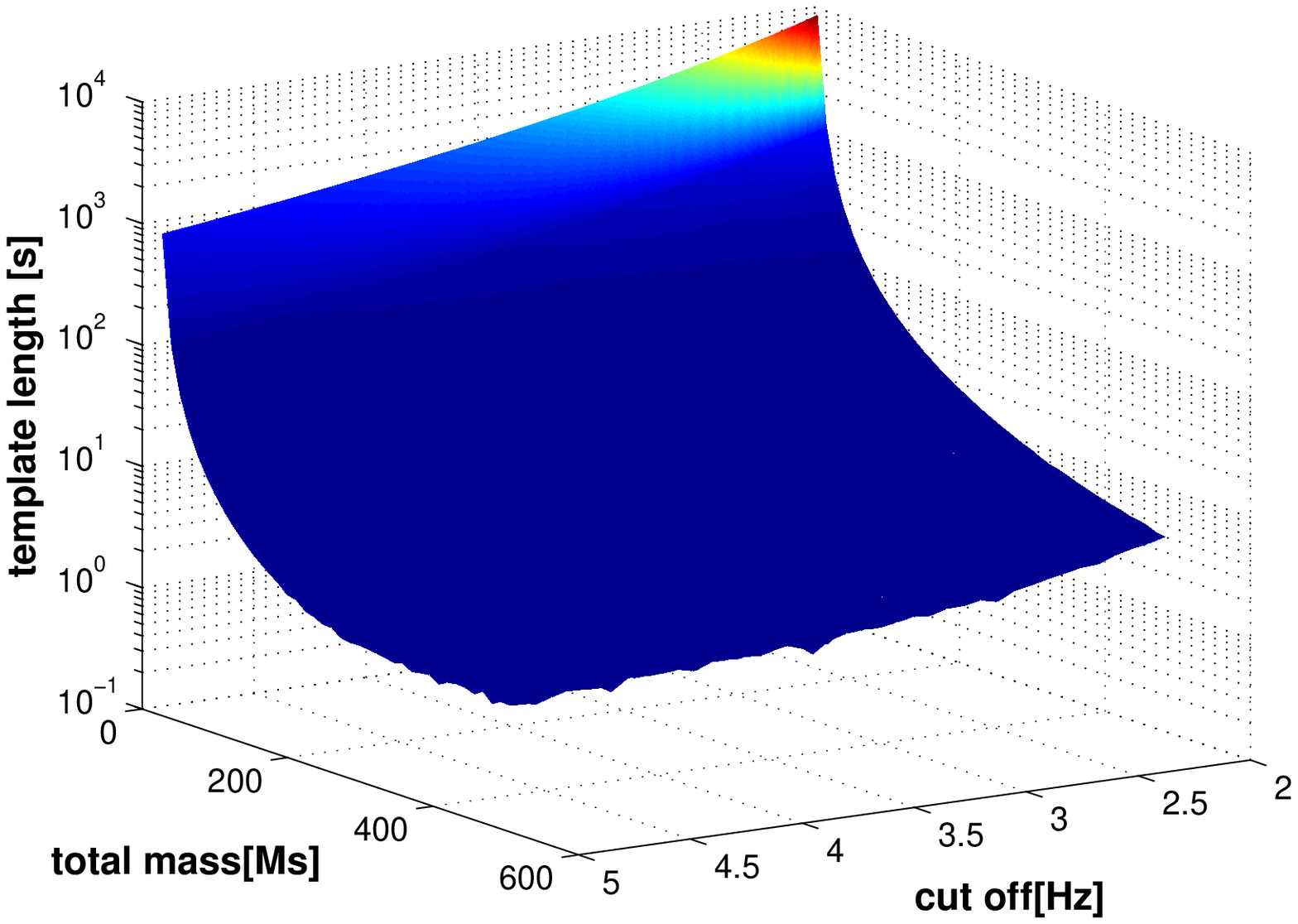}
    }
  }

  \vspace{5pt}
    \hbox{\hspace{1.in} (a) \hspace{1.9in} (b)}   
  \vspace{7pt}

  \centerline{\hbox{\hspace{0.0in}
    \epsfxsize=2.7in
    \epsffile{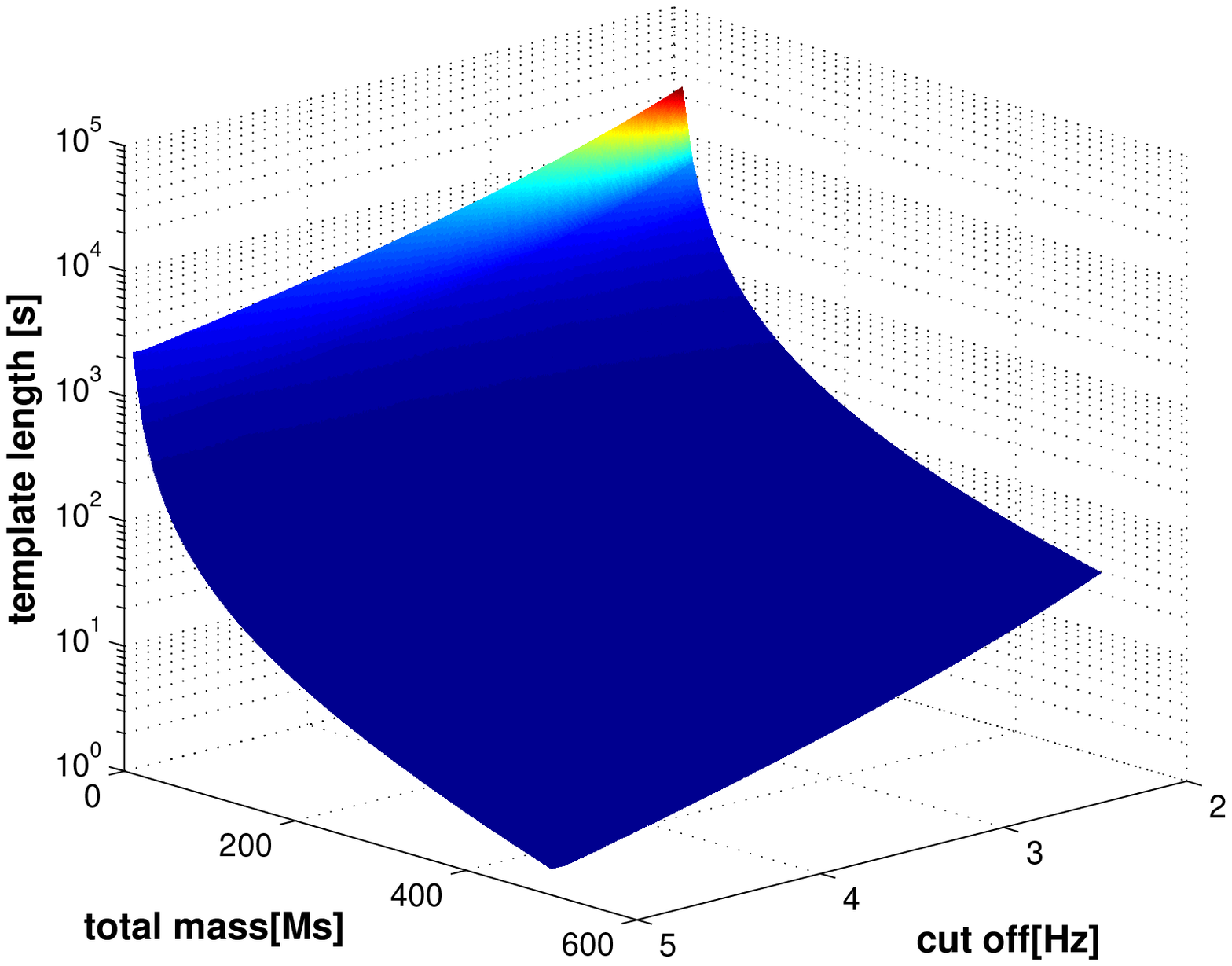}
    \hspace{0.05cm}
    \epsfxsize=2.7in
    \epsffile{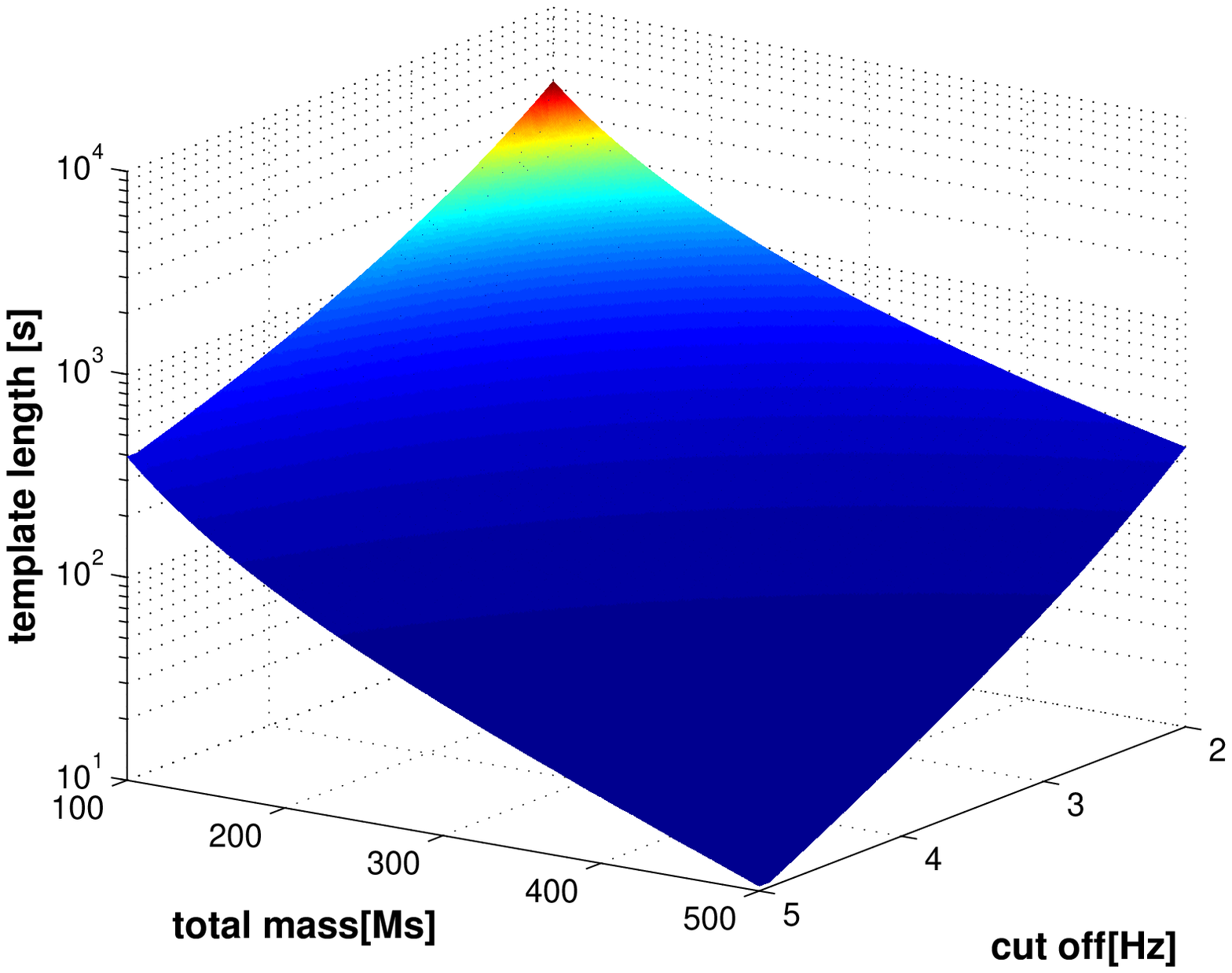}
    }
  }

  \vspace{5pt}
  \hbox{\hspace{1.in} (c) \hspace{1.9in} (d)} 
  \vspace{5pt}

  \caption{Signal duration in the ET detector based on total mass, mass ratio and lower frequency cut-off of the detector.  The systems represent equal mass binaries with total masses of $1\leq m/M_{\odot}\leq10$ (a), equal mass binaries with total masses between $10\leq m/M_{\odot}\leq 500$ (b), binaries with a total mass between $10\leq m/M_{\odot}\leq500$ for mass ratios of 10 (c) and  binaries with a total mass between $100\leq m/M_{\odot}\leq500$ for mass ratios of 100 (d).}
  \label{fig:durations}

\end{figure}

\section{Sources for ET. \label{ets}}
In this section we review some of the sources expected to be visible in the ET detector.  We also look at the problems that some of these sources may present to any future data analysis strategies
\subsection{Source duration in ET.}
The main sources of interest for ET will be the inspiral of stellar mass compact objects such as neutron star - neutron star (NS-NS), neutron star - black hole (NS-BH) and black hole - black hole (BH-BH) binaries.  While these are common sources to the current ground based detectors, they will be viewed in a different way by ET and will in fact present their own special problems.  Depending on the the low frequency cut-off of the detector, these sources will be very long lived in the detector.  In Figure~\ref{fig:durations} we plot the duration signals are visible in the detector for four different cases.

Going from left to right, and top to bottom, the durations are plotted for equal mass binaries with total masses of $1\leq m/M_{\odot}\leq10$ (a), equal mass binaries with total masses between $10\leq m/M_{\odot}\leq 500$ (b), binaries with a total mass between $10\leq m/M_{\odot}\leq500$ for mass ratios of 10 (c) and  binaries with a total mass between $100\leq m/M_{\odot}\leq500$ for mass ratios of 100 (d).  We can see that, depending on the lower frequency cut-off of ET, the above sources will be very long lived in the detector.  For example, assuming a low frequency cut-off of 2 Hz, a $(10, 10)\,M_{\odot}$ binary will last about 45 minutes in ET, as opposed to 38 seconds for a second generation detector with a low frequency cut-off of 10 Hz and 0.9 seconds for a first generation LIGO detector with a low frequency cut-off of 40 Hz.  For a $(1.4, 1.4)\,M_{\odot}$ system, the durations are approximately 20 hours for ET,  16 minutes for a second generation detector and approximately 25 seconds for initial LIGO.

While it is clear that it will be advantageous to have longer lasting signals, ET will face new issues.  It is possible that at low frequencies, the large number of NS-NS binaries may form a confusion background making it difficult to resolve particular sources.  This is a problem that is well known in terms of LISA data analysis with white dwarf binaries.  

\subsection{LISA - ET sources}
As well as the sources defined above, there may also be what can be termed as LISA-ET sources.  Assuming a co-existing LISA-type spaced based detector, it may be possible to have sources that are visible firstly in the space detector (for example during the inspiral phase) and later in ET (during the merger and ringdown).  For example, if we consider the inspiral of two intermediate mass black holes (IMBHs) in a nearby globular cluster, it has been shown that the inspiral of such sources would be visible in LISA~\cite{seps,pl}.  Assuming equal masses, a total binary mass of 800 $M_{\odot}$ and an initial GW frequency of 5 mHz, the binary would be visible in the LISA band for approximately a year.  Once the binary then entered the ET band, assuming a low frequency cut-off of 2 Hz, it would take the system about 5 seconds to reach the last stable circular orbit at 5.5 Hz.  This system would have a merger frequency at approximately 20 Hz, with the ringdown phase pushing the observable duration even further.

Another observable system in both LISA and ET could be an intermediate mass ratio inspirals (IMRIs)~\cite{jg}.  While the astrophysical evidence for such systems is scant, they still require consideration for a data analysis strategy.  If for example we take a system with a total mass of 1000 $M_{\odot}$, with a mass ratio of $\sim1000$, and again assume a low frequency cut-off of 2 Hz, the IMRI would be visible in ET for a little over 10 minutes.  

As well as inspiralling binaries, it was shown that more exotic objects such as the GW emission from kinks and cusps of cosmic strings may also be observable coincidently in both LISA and LIGO~\cite{key}.  Such objects would also be visible in ET. 

\subsection{Increased parameter space}
With the increase in source type, we also acquire an increase in the complexity of the data analysis problem due to the increased parameter search space.  As the sources are spending so much longer in the band of the detector, it should be possible to carry out advanced parameter estimation for parameters such as the masses, sky position and luminosity distance.  However, with ET we would also expect to be able to discern the inclination of the orbital plane, the eccentricity of the binary and the spins of the component bodies.  In most cases this will increase the  number of parameters of between 15 and 17, depending on the source.

The increase in the dimensionality of the parameter space brings with it additional computational cost.  For a standard template bank, the number of search templates needed increases geometrically with dimension.  It is therefore not hard to see that computational costs can very quickly get out of hand as we increase the number of search dimensions.

\subsection{Higher waveform harmonics, merger \& ringdown}
In the last few years a number of studies have been carried out in the LISA framework on the effect of higher harmonic corrections to the waveforms and parameter estimation~\cite{pc,ts,arun1,arun2}.  The overwhelming result of these studies was that higher harmonics are an essential ingredient to any data analysis strategy for inspiralling compact binaries.  The higher harmonic corrections effect the data analysis in number of ways.  The first is that the extra harmonics help break correlations between the parameters.  The second is an increase in the precision of the parameter extraction.  

The extra precision can be reasoned with the argument of variable counting : If we consider a binary of two non-spinning black holes, we can describe the system with nine parameters.  Then, if a binary is visible long enough in the detector, we can measure the frequency and the first and second derivatives of the frequency with precision.  This provides us with three variables for the nine coefficients.  We should also be able to measure the time dependent amplitude and phase of the detector response, providing another two variables.  Thus in total we have five variables for nine parameters.  Now, if we include two extra harmonics, we end up with fifteen variables for the nine parameters.  Thus not only do we have an over-described parameter set, but along with the breaking of parameter correlations, we now have the ability to carry out extremely accurate parameter estimation. 

Another way in which the higher harmonics can impact the data analysis is by allowing higher mass systems to be visible in the detector.  While the dominant harmonic component of a particular source may be outside the bandwidth of ET, it is possible that one of the higher harmonics will be visible in the detector.  If the signal to noise in these extra harmonics are strong enough, it would allow us to detect and carry out parameter estimation for sources that would otherwise be lost to the detector.

As well as the inclusion of higher harmonics, an inspiral search should also include the merger and ringdown phases of the waveform.  While maybe less important for systems spending a long time in the detector, the inclusion of merger and ringdown will
be important for heavier sources that are terminating at lower frequencies in the detector.  To reuse an example given above, the inspiral of two IMBHs of $(400,400)\,M_{\odot}$ has an LSO frequency of about 5 Hz, whereas the merger frequency of such a source is at approximately 20 Hz.  We can see that including both the merger and ringdown for this type of source could greatly enhance parameter estimation.

\section{Usefulness of LISA oriented algorithms \label{lisaal}}
One of the consequences of the increase in the parameter search space and the need to include more complicated waveform models is that current ground based search techniques such as template banks may become computationally unwieldy for ET.  In the last five years or so, great progress has been made in the LISA community on the development of algorithms capable of searching in high dimensional spaces for white dwarf - white dwarf galactic binaries, the inspiral of comparable mass massive and supermassive black hole binaries (MBHBs/SMBHBs) and extreme mass ratio inspirals (EMRIs).  Due to the fact that we may have large source numbers and high signal to noise ratios, it may be that current LISA search algorithms could be easily adapted and applied to ET data analysis.  In the next few subsections we present some of the tried and tested LISA algorithms.

\subsection{Markov Chain Monte Carlo based methods}
The Markov Chain Monte Carlo (MCMC) is a stochastic method which is ideal for searching through high dimensional spaces and mapping out posterior density functions.  It works by constructing a chain of solution points in parameter space drawn from a proposal distribution that we believe to be close to the target density we are trying to model.  If the chain is run long enough then we are guaranteed to eventually map out the target density.  There are a number of types of MCMC, such as a Gibbs algorithm (where we update one parameter at a time) or a Metropolis-Hastings algorithm (where we update all parameters at the same time). 

While MCMC methods have been used in the ground based community~\cite{ch1,ch2,ch3,ch4}, a variant of the MCMC called the Metropolis-Hastings Monte Carlo (MHMC)~\cite{cp1,cp2,cp3} was developed to search for SMBHBs in LISA.  It has since been applied with great success to searches for SMBHBs and and more recently EMRIs in blind Mock LISA data analysis challenges (MLDCs)~\cite{mldc}.  Using a number of annealing schemes, as well as directed jump proposals, these algorithms are both fast and accurate.  A variation called Parallel Tempered MHMC (PTMCMC) is also being used in searches for EMRIs, spinning SMBHBs and cosmic strings.   The PTMCMC uses a number of different chains at different temperatures to search the parameter space.  By allowing communication between chains via a genetic algorithm, it is possible to quickly converge on a solution.

The main advantage in using MHMC methods is that while the number of templates needed in a template bank scales geometrically with the dimensionality of the search space, the MHMC method scales linearly.  In searches for SMBHBs, the MHMC algorithm found the sources, sometimes using less than 5000 templates over the entire parameter priors, in a five dimensional search space.

\subsection{Dealing with source confusion}
One of the great successes of MCMC based codes is the Block Annealed Metropolis-Hastings or BAM algorithm~\cite{cc1,cc2}.  This algorithm was designed to extract individual white dwarf galactic binaries from a galactic foreground consisting of approximately 30 million binaries.  The theoretical prediction is that about 25,000 of these binaries are resolvable, with the rest remaining as an additional type of noise called confusion noise.  In a blind MLDCs, the BAM algorithm resolved approximately 20,000 individual binaries, almost approaching the theoretical limit.  If indeed there will be a NS-NS confusion background in ET, it is conceivable that an algorithm such as the BAM algorithm could be used to extract the sources.

\subsection{Multi-modal solution methods}
One of the main problems in GW data analysis is that we never know a priori just how many signals there are.  It is always useful to have algorithms that can find multiple solutions (if there are many sources) or multiple modes (if the number of sources is small, but the solution is degenerate).  One of the first algorithms of this kind to be applied to GW data analysis was based on a technique called Nested Sampling~\cite{skilling,vecchio}.  This algorithm works by using a number of live solution points in the parameter space to climb towards the peaks of highest likelihood.

One of the issues with Nested Sampling, was that as one approaches the true solution it becomes harder to find a better solution just by randomly sampling from inside the parameter priors.  Two improvements were made to the standard Nested Sampling in the context of GWs.  The first was a Hybrid Evolutionary Algorithm~\cite{hea} which used a mixture of Nested Sampling, MHMC and elements of evolutionary computing, such as birth, death, altruism etc.  The second was an algorithm called MultiNest~\cite{mn} that was originally applied to particle physics and cosmology, but has since also been applied to GW data analysis~\cite{mngw}.  These algorithms are again very fast and very accurate.

Another form of evolutionary algorithm, called a genetic algorithm, was first developed for use with galactic binaries~\cite{ga1} and has since been used in searches for spinning massive black holes~\cite{ga2}.

\section{Computational issues for ET \label{etcc}}
\subsection{Some future costs}
In this section we present a plausible forecast on computational issues for the ET detector. In particular, the inspiral case is take into account. The reason of this choice is related to the fact that we expect a high number of such events. Furthermore,  the fundamental detection procedure involved in this analysis is the matched filter, which as we have already stated, is computationally expensive.   However, due to the prominence of the method in both LIGO/VIRGO and LISA data analysis, we can expect that in the future of inspiral signal detection, the matched filter will again be used, but with some important evolution due to the information acquired from the first direct observation and more complex waveforms. An important first consequence, due to the high computing power that should be available in future architectures and evolution in the waveform modelling, will be the possibility of following the GW signal from inspiral through to merger and ring-down. 

As regards the burst detection evolution, we can expect that while some suboptimal methods for unmodeled sources will be kept and evolved, other methods probably based on pattern recognition or more sophisticated procedure will be introduced. Simultaneously, we can expect that improvements in simulation and astrophysical theory will allow us to better model different sources and signals, thus permitting the application of matched filtering, of which the computational requirements are given by the dimensions and size of the parameter space.

In general, it is important to remember that the main goal of ET is the observation of sources as a GW telescope. This implies the necessity to introduce a completely new set of tools. Firstly, with the purpose of conducting accurate parameter estimation, and secondly a multi-messenger analysis which involves a completely new set of procedures that should allow an optimal signal detection.

\subsubsection{Inspiral case} 
In this subsection we present some rough estimations for the future computational cost of an ET inspiral search, using as a reference, the actual analysis procedure for compact binaries. As shown in Figure \ref{fig:durations}, the length of the GW inspiral signal observable by ET covers a wide range of values. Starting the analysis at 2 Hz, we pass from more than 1 day of observation for low masses, to a few seconds or less for the highest ones. it is clear that this population of signals cannot be handled with a single approach, but will require some hierarchical method.

\begin{figure}[t]
\centering
\epsfig{file=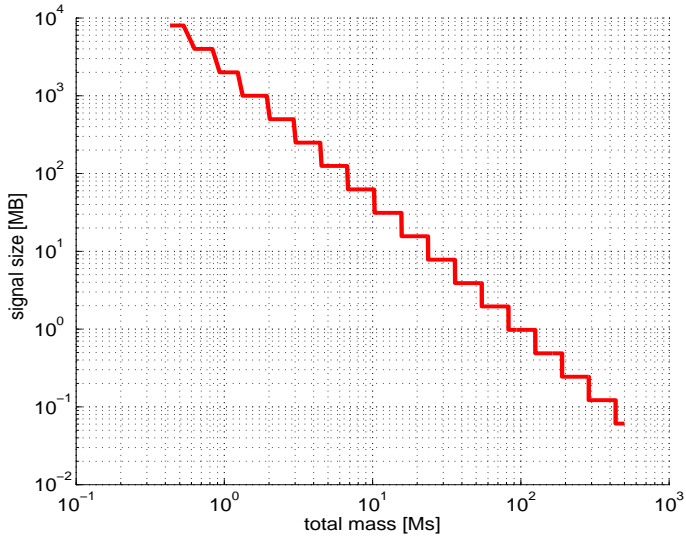, width=4in, height=3.in}
\caption{A plot of template size in MB versus total system mass assuming that all templates have the same sampling rate of 4 kHz.}
\label{fig:sizeinspiral}
\end{figure}

\begin{figure}[t]
\centering
\epsfig{file=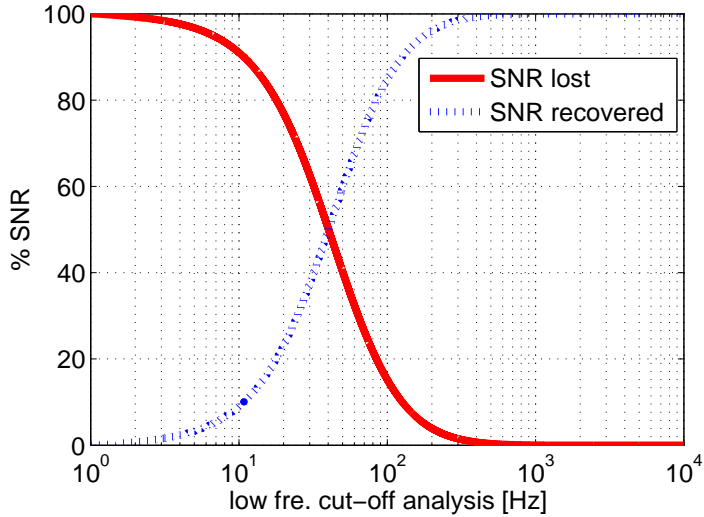, width=4in, height=3.in}
\caption{A plot of the percentage of SNR both lost and recovered as a function of the low frequency cut-off used in the detection analysis.}
\label{fig:etcumulative}
\end{figure}

So, the first consideration when it comes to computational issues is the generation procedure and archiving needed to handle and process such very long signals. Our introductory analysis is based on a naive sub-optimal approach, i.e. generating full templates with the same sampling frequencies regardless of total mass or the highest frequency content of the template.  In this case , if we consider single-precision storage and a 4 kHz sampling rate, template sizes change by an order of magnitude, starting from less than 1MB for the highest masses up to many GB for lighter binary systems.  These results are shown in Figure~\ref{fig:sizeinspiral}.
Such long templates produce a devastating effect on both the generation time and archiving procedure.  The template length also effects the filtering process due to the fact that, as already stated, the matched filtering algorithm has a dominant computational complexity based on the FFT that is $O(Nlog(N))$, where $N$ is the number of samples composing the signal. 
It is obvious that we have to introduce some tricks to balance the analysis. A possible solution can be achieved by taking some considerations into account. A first observation is regarding the cumulative SNR curve based on the ET sensitivity B curve\cite{etb}, as  shown in Figure \ref{fig:etcumulative}.
These curves show a rough estimation of the percentage SNR lost (solid line) and the percentage SNR recovered (dashed line) with respect to the low frequency cut-off at which the analysis starts. From these curves we learn that we can arbitrarily identify a sub-boundary $[f_{start}-f_{stop}]$, losing a few percent in signal-to-noise for detection purposes.

\begin{figure}[t]
\centering
\epsfig{file=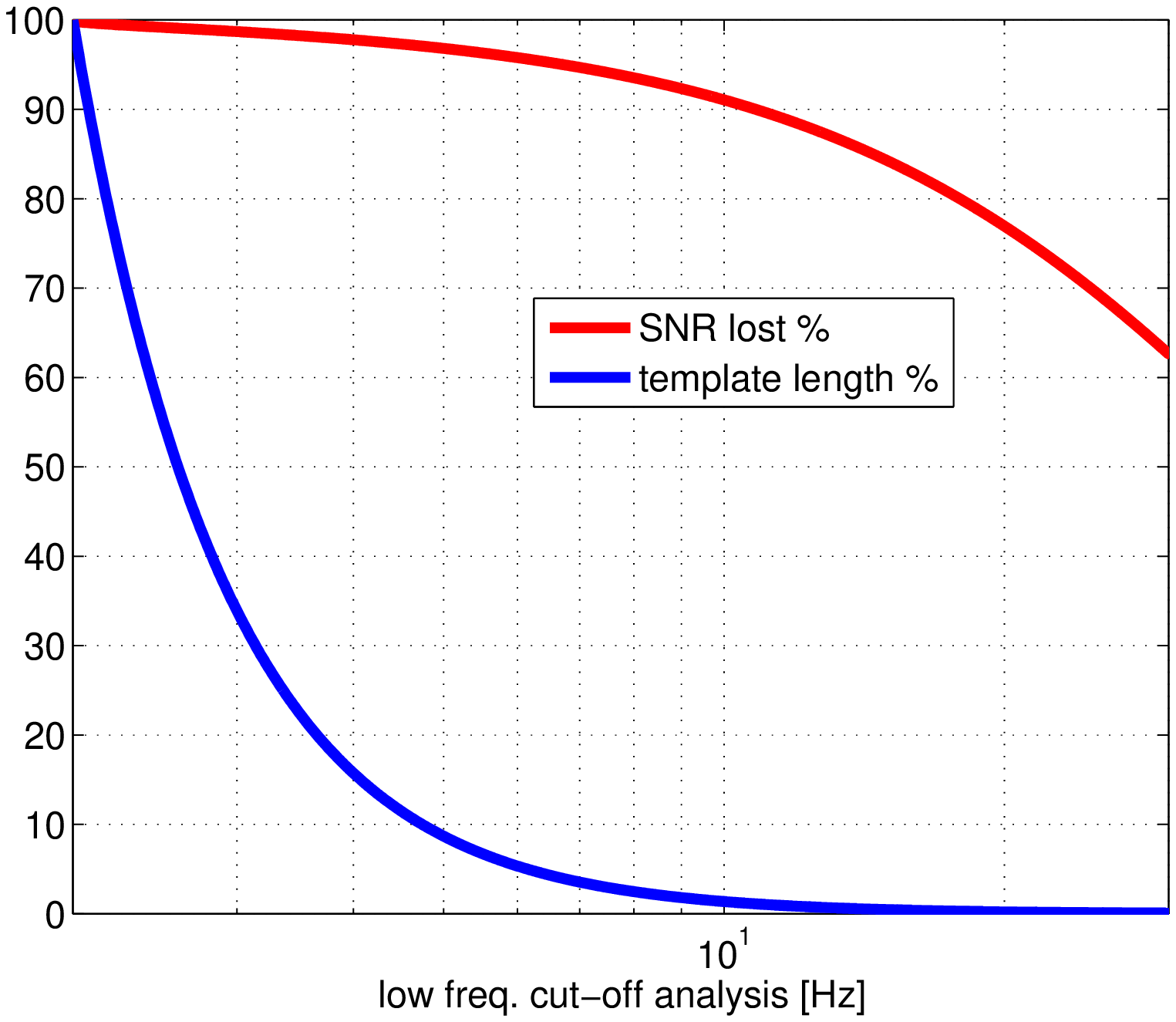, width=4in, height=3.in}
\caption{A plot of the percentage of lost SNR and residual template length as a function of low frequency cut-off.}
\label{fig:ET-snr-length}
\end{figure}

Starting with this in mind, we can proceed with a second observation related to the time/frequency characterization of an inspiral signal. In fact while the system evolves towards coalescence, it has a frequency which develops exponentially. This means, for example, that the binary system takes more time to move from 2 Hz to 3 Hz, than from 10 Hz to 11 Hz. If we consider a binary system composed of two NS with identical individual masses of $1\,M_{\odot}$, it is possible to estimate the following data:  the system spends $91\%$ of the time (more than 1 day in this case) at very low frequency, where a sampling frequency of 5 Hz could be enough for signal detection and with a contribution to SNR less than $3\%$. This concept is demonstrated in Figure \ref{fig:ET-snr-length}, where we report the percentage of lost SNR and the percentage of the residual template length as a function of the low frequency cut-off. Using our approximation, at 4 Hz we observe a template length reducion of $90\%$ but with a loss in SNR of a few percent. 
\begin{table}[t]
\begin{center}
\begin{tabular}{ c | c ||r | c }
from[Hz] & to [Hz] & time[s] & \% of total duration  \\
\hline
2 & 3 & 85000 s 	& 66\%		\\
2 & 5 & 117000 s & 91\%	\\
5 & 10 & 9500 s 	& 7.5\%	\\
10 & 1kHz & 1776 s & 1.5\%\\
\hline
\end{tabular}
\end{center}
\caption{This table compares the duration and percentage of the total duration a $(1,1)\,M_{\odot}$ NS-NS binary spends in various parts of the frequency band between 2 Hz and 1 kHz.}
\label{tab:snr40mle}
\end{table}

This type of waveform generation would allow us to gain orders of magnitude on the template generation time, the number of template processable per second, and data archived, with respect to the first naive approach of identical sampling.  In fact, still considering the $(1,1)\,M_{\odot}$ system, instead of requiring half a GB of memory, we can, for example, generate templates using a  multi-sample-rate format as displayed in Table~\ref{tab:msr}.  This produces an occupation space of less than 13 MB versus half a GB, with a gain in matched filter performance of two orders of magnitude.
\begin{table}[t]
\begin{center}
\begin{tabular}{ c | c ||r | r | r}
from[Hz] & to [Hz] & time[s] & sample rate[Hz] & size [MB]\\
\hline
2 & 5 & 117000 s & 10Hz	& 4MB\\
5 & 10& 9500 s & 20Hz	& 1MB\\
10& 1kHz & 1776 s & 4kHz	& 8MB\\
\hline
total: &&&& 13MB \\
\hline
\end{tabular}
\end{center}
\caption{This table illustrates the reduced cost of a multi-sample-rate template.  By sampling to different frequencies at different sampling rates, the total size of a $(1,1)\,M_{\odot}$ NS-NS system can be reduced from 0.5 GB to 13 MB.}
\label{tab:msr}
\end{table}
So, a possible approach for ET data analysis as a GW observatory, could be to define something like a \textit{reverse followup analysis}, dividing the investigative process in various steps. The first step, for example, could be to try and catch the inspiral process from where it is more important for the detection point of view, and where it is possible to use a faster and possibly sub-optimal approach. In this targetting phase we have to take into account the fact that in the ET era, the template will be composed of inspiral,merging and ringdown phases.  The next step performs a reverse followup of the events with a multi-sample rate analysis, introducing more accuracy in following the events and switching into observation rather than detection mode.

Last but not least, an important estimation concerns the number of templates needed by a bank of filters for detection and signal extraction purposes. Here we present some preliminary and very rough results obtained using the ET-B sensitivity curve and an inspiral library\cite{inspiral}, assuming a $95\%$ minimal match.  
\\
\begin{table}[t]
\begin{center}
\begin{tabular}{ c | c || r }
low freq. cut off[Hz] & individual masses$[M_{\odot}]$ & \# templates \\
\hline
10 & 1 - 3 & 40000 \\
5 &  1 - 3 & 80000 \\
2 & 1 - 3 & 150000 \\
2 & 1 - 500 & 1000000 \\
\hline
\end{tabular}
\end{center}
\caption{This table shows the approximate number of templates needed for an ET inspiral search assuming a minimal match of 95\% and different lower frequency cut-offs and individual mass ranges for the analysis.}
\label{tab:snr40mle}
\end{table}
\\

\section{Summary and Discussion \label{summary}}
We have discussed some of the data analysis issues for the planned third generation GW detector, the Einstein Telescope.  We have shown that ET should be able to detect the inspirals, mergers and ringdowns of compact binaries, intermediate mass black hole binaries and intermediate mass ratio inspirals.  The systems will be long lived in the detector, thus allowing the community to carry out a detailed parameter estimation of the sources.  There is also the possibility that neutron star binaries at low frequencies could be a source of confusion noise make resolution of individual sources difficult.  In this respect, algorithms that have been developed for LISA data analysis may be of use for ET.  The richness in the number of sources, as well as the increased parameter search space also infers a number of computational issues.  The longer waveforms require more computing power and hence increased storage.   In this article, we presented a rough estimate of the number of templates needed for an inspiral search for ET, the computational cost per template and how this cost can be reduced by using a multi-sampling-rate waveform generation.

\section*{Acknowledgments}


\end{document}